\documentclass[aps, prb, twocolumn, superscriptaddress, longbibliography]{revtex4-2}
\usepackage[colorlinks, linkcolor=blue, anchorcolor=blue, citecolor=blue]{hyperref}
\usepackage{amsmath}
\usepackage{graphicx}
\usepackage{braket}
\usepackage{multirow}
\usepackage{color}
\usepackage{soul}
\usepackage{diagbox}
\usepackage{rotating}

\usepackage{ulem}
\begin{document}

\title{Hardening of Ni-O bond-stretching phonons in LaNiO$_2$}
\author{Yilin Wang}\email{yilinwang@ustc.edu.cn}
\affiliation{School of Emerging Technology, University of Science and Technology of China, Hefei 230026, China}
\affiliation{Hefei National Laboratory, University of Science and Technology of China, Hefei 230088, China}
\affiliation{New Cornerstone Science Laboratory, University of Science and Technology of China, Hefei, 230026, China}

\date{\today}

\begin{abstract}
    We demonstrate that dynamical electron correlation and fluctuating local magnetic moments are crucial for the phonon spectra of the infinite-layer nickelate superconductor, LaNiO$_2$, using DFT plus dynamical mean-field theory (DFT+DMFT) calculations. We find significant hardening of optical Ni-O bond-stretching phonons when going from non-magnetic to paramagnetic state, and increasing Coulomb interaction will make them even harder. The electron correlation is found to be sensitive to the Ni-O bond-stretching distortions, indicating strong interplay between electron correlation and lattice. 
    We find that the strong local electron correlation will not favor charge orders that couple to the Ni-O bond-stretching phonons, in support of the recent experiment that a $3a_0$ charge order is absent in the infinite-layer nickelates. Our results emphasize that the effects of local magnetic fluctuations should be fully taken into account when describing the lattice dynamics of the infinite-layer nickelates without long-range magnetic orders, and also provide evidence to ascribe the kink observed in the recent angle-resolved photoemission experiment to possible strong electron coupling to the Ni-O bond-stretching phonons.
\end{abstract} 

\maketitle

\section{Introduction} 
The discovery of superconductivity in the infinite-layer nickelate Nd$_{1-x}$Sr$_x$NiO$_2$~\cite{Danfeng:2019} brings the study of unconventional superconductivity into the nickel age~\cite{Sawatzky:2019,Norman_trend:2020,Pickett:2021}. 
A comparison of the phase diagram of the infinite-layer nickelates to copper oxide superconductors is crucial to understand the superconducting mechanism since they share similar $3d^9$ configuration~\cite{Anisimov:1999}. Extensive investigations have shown both similarities and differences between the infinite-layer nickelates and cuprates~\cite{Pickett:2004,Hirofumi:2019,GeorgeSawatzky:2019,Peiheng:2019,Nomura:2019,Hepting:2020,Dangfeng:2020,Botana:2020,Ariando:2020,ZhangHu:2020,ZhangYahui:2020,Xianxin:2020,MiYong:2020,Rossitza:2020,GoodgeBerit:2021,Atsushi:2021,JiaChunjing:2021,KuWei:2021,GeorgeSawatzky:2023,HanXiangru:2023,SiLiang:2023,GuangMing:2020,Olevano:2020,Kutepov:2021,Louie:2022,Meier:2024}. In particular, strong electron correlations~\cite{YilinWang:2020,Karp:2020,SiLiang:2020,Ryee:2020,Leonov:2020,Kitatani:2020,Lechermann:2020,Werner:2020,Petocchi:2020,Kangchangjong:2021,Sangkook:2021,Craco:2022,Kang2023} and magnetic fluctuations~\cite{GuYuhao:2020,Leonov:2020,LiuZhao:2020,Lechermann:2021mag,YuWeiqiang:2021,WuTao:2021,WanXiangang:2021,Slobodchikov:2022,Andreas:2022,Hepting:2022,Sahinovic:2023,ZhuJianxin:2023,Pascut2023,Devereaux:2024,ZhangRuiqi:2024} are revealed in the infinite-layer nickelates, which are crucial for unconventional superconductivity~\cite{LunHui:2019,AndrewMillis:2022,Lechermann:2022,Held:2023,DiCataldo:2024}. The strong magnetic fluctuation is one of the reasons leading to the absence of long-range magnetic order~\cite{Hayward:1999}. Even so, dispersive magnetic excitations are still observed by resonant inelastic x-ray scattering experiments~\cite{LeeWeisheng:2021,ZhouKejin:2022,Preziosi:2022,Hepting:2024,ZhuZhihai:2024,Rossi:2024}, similar to the doped cuprates~\cite{Kazuyoshi:2006,LeTacon:2011,MarkDean:2013,Minola:2015,PengYY:2018}. Recently, two angle-resolved photoemission spectroscopy (ARPES) experiments~\cite{FengDonglai:2024,NieYuefeng:2024} have successfully observed cuprate-like electronic structures in LaNiO$_2$. 

Studies show that phonons may play important roles in the superconductivity~\cite{Devereaux:2021,Devereaux:2023} and charge order~\cite{Ghiringhelli:2012,Chang:2012,Comin:2016,MiaoHu:2018,Huang:2021} physics in cuprates. In particular, the optical Cu-O bond-stretching phonons are essential~\cite{McQueeney:1999,Reznik:2010,Reznik:2012,Reznik:2014,Reznik:2020}. Similar to cuprates, signatures of possible strong electron-phonon coupling and charge order have been observed experimentally in the infinite-layer nickelates. Firstly, a low-energy kink around 70$\sim$100 meV has been observed in the electronic structure of LaNiO$_2$ by ARPES experiment~\cite{NieYuefeng:2024}. Such kinks are usually ascribed to strong electron coupling to phonons or spin excitations. Secondly, a stripe-like $3a_0$ charge order is observed in the infinite-layer nickelates~\cite{ZhouKejin:2022,Preziosi:2022,LeeWeisheng:2022} and different driving forces have been proposed~\cite{ChenHanghui:2023,HuangBing:2023,JiaChunjing:2023,Stepanov:2024,DengHuiXiong:2024}. It suggests that this order may couple to the Ni-O bond-stretching phonon around $Q=(1/3, 0, 0)$ and lead to phonon softening~\cite{LeeWeisheng:2022}. However, a recent transmission electron microscopy (TEM) experiment demonstrates that this charge order is not intrinsic but induced by the ordering of excess apical oxygens~\cite{ShenKM:2024}, which is supported by further x-ray scattering experiment~\cite{Hepting:2024}. This raises a question as to whether there exists intrinsic (potential) charge order in infinite-layer nickelates and how the strong electron correlation affects it. It has shown that strong electron correlation is crucial for correctly describing the Cu-O bond-stretching phonons in cuprates~\cite{Reznik:2021}. In the infinite-layer nickelates, how the strong dynamical electron correlation and fluctuating local magnetic moments affect their phonon dynamics has not been established. For antiferromagnetically ordered systems such as cuprates and UO$_2$, calculations based on DFT+$U$ method in antiferromagnetic states can well describe their phonon spectra~\cite{Reznik:2021,Shuxiang:2022}. In contrast, there are no long-range magnetic orders but strong fluctuation of local magnetic moments in LaNiO$_2$. Thus, it requires phonon calculations in paramagnetic (PM) state, taking account of the strong dynamical electron correlations.

\begin{figure*}
    \centering
    \includegraphics[width=1.0\textwidth]{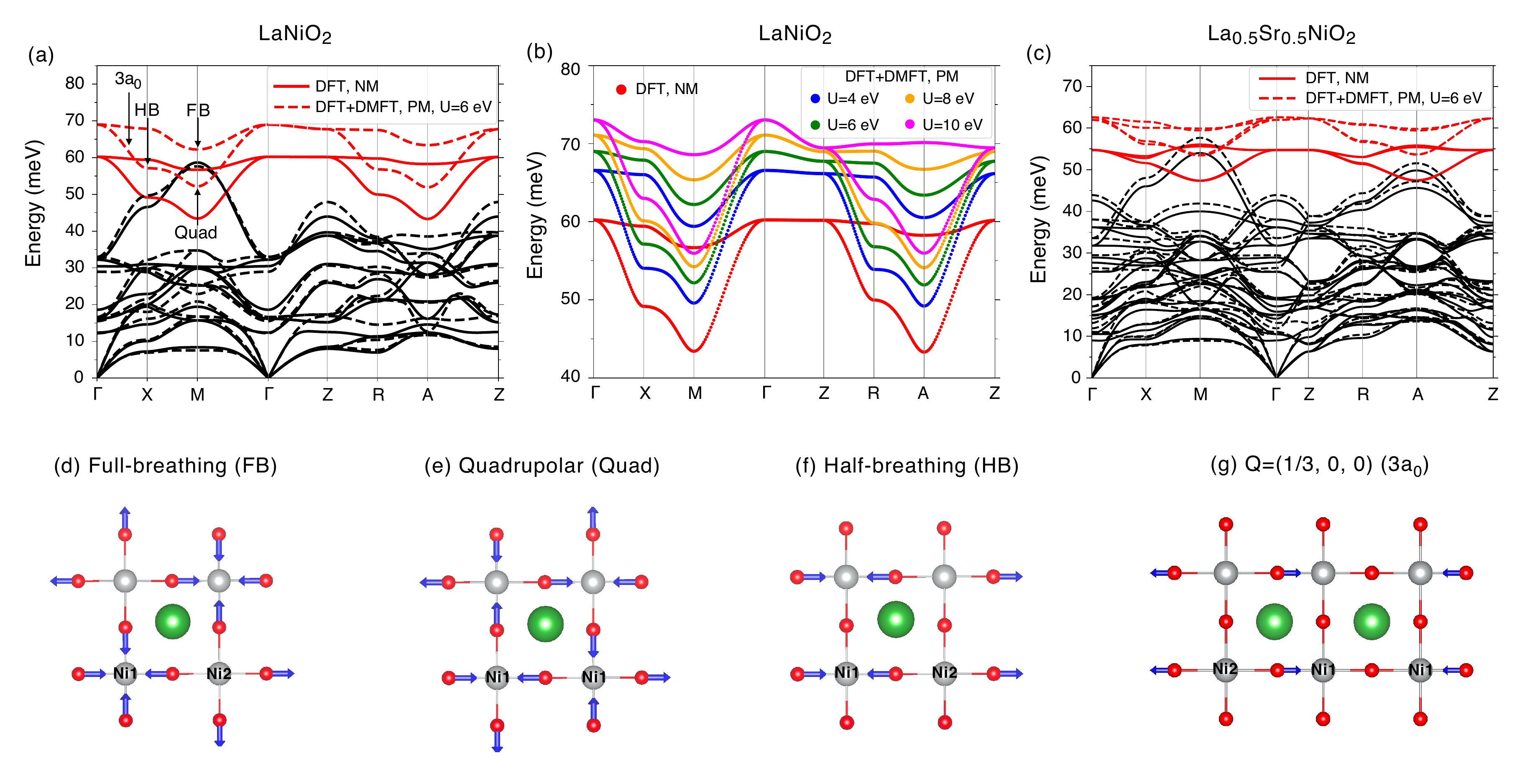}
    \caption{Hardening of Ni-O bond-stretching phonons. (a) Phonon spectra of LaNiO$_2$ calculated by DFT in the NM state (solid curves) and by DFT+DMFT in the PM state at $U=6$ eV (dashed curves), respectively. The optical Ni-O bond-stretching phonons are highlighted by the red color. (b) DFT+DMFT calculated Ni-O bond-stretching phonons as function of Hubbard $U$. (c) DFT and DFT+DMFT calculated phonon spectra of 50\% Sr-doped compound La$_{0.5}$Sr$_{0.5}$NiO$_2$. (d)-(g) Illustration of atomic vibrations of four Ni-O bond-stretching phonon modes, including full-breathing (FB), quadrupolar (Quad), half-breathing (HB) and $Q=(1/3, 0, 0)$ ($3a_0$), as indicated by the black arrows in (a). There are two non-equivalent Ni sites for FB, HB and $3a_0$.}
    \label{fig:dmft_phonon}
\end{figure*}

In this work, we study the effects of dynamical electron correlation on the lattice dynamics of the infinite-layer nickelate superconductor, LaNiO$_2$, by computing its phonon spectra directly in the PM state based on DFT plus dynamical mean-field theory (DFT+DMFT) method~\cite{Georges:1996,lichtenstein:2001,kotliar:2006}. We find significant hardening of optical Ni-O bond-stretching phonons when going from non-magnetic (NM) to PM state, and increasing Coulomb interaction will make them even harder. We demonstrate that such hardening effects are induced by the dynamical electron correlation and fluctuating local magnetic moments. We also find that the electron correlation is very sensitive to Ni-O bond-stretching distortions. Furthermore, we find that the strong local electron correlation will not favor charge orders that couple to the Ni-O bond-stretching phonons, in support of the recent experimental finding that the $3a_0$ charge order is absent in the infinite-layer nickelates~\cite{ShenKM:2024,Hepting:2024}. Our results emphasize that the effects of local magnetic fluctuations should be fully taken into account when describing the lattice dynamics of the infinite-layer nickelates without long-range magnetic orders.

\section{Methods} 
Fully charge-self-consistent DFT+DMFT calculations are performed using the DFT+eDMFT code developed by Haule \textit{et al.}~\cite{Haule:2010,Haule:2015free} based on the WIEN2K package~\cite{Blaha:2020}. This code can compute very accurate forces as long as providing accurate self-energy~\cite{Haule:2016force}, so the phonon spectra can be computed in PM state by the finite-displacement method with the aid of the \textit{phonopy} package~\cite{KristjanHaule:2020,phonopy-JPCM,phonopy-JPSJ}. A large hybridization energy window from -10 to 10 eV is included and all the five $d$-orbitals are considered as correlated ones in the DMFT calculations. An ``exact'' double-counting scheme is chosen~\cite{Haule:2015}. We also test other double-counting schemes and they will not change the conclusion. The continuous time quantum Monte Carlo~\cite{Gull:2011} is used as impurity solver. We use the experimental crystal structure of LaNiO$_2$ with lattice parameters $a=b=3.966$ \AA\: and $c=3.376$ \AA~\cite{Levitz:1983}. We simulate 50\% Sr-doping in a $1\times1\times2$ supercell (see Fig. S1 in the Supplementary Materials~\cite{suppl}). The DFT+$U$ calculations are performed using the scheme introduced by Anisimov et al.~\cite{Anisimov:1993}, with an approximate correction for the self-interaction correction (SIC) in the WIEN2K package~\cite{Blaha:2020}. The Hubbard $U$ and Hund's coupling $J_H$ are explicitly included.
More computational details are presented in the Supplementary Materials~\cite{suppl}.


\begin{figure*}
    \centering
    \includegraphics[width=1.0\textwidth]{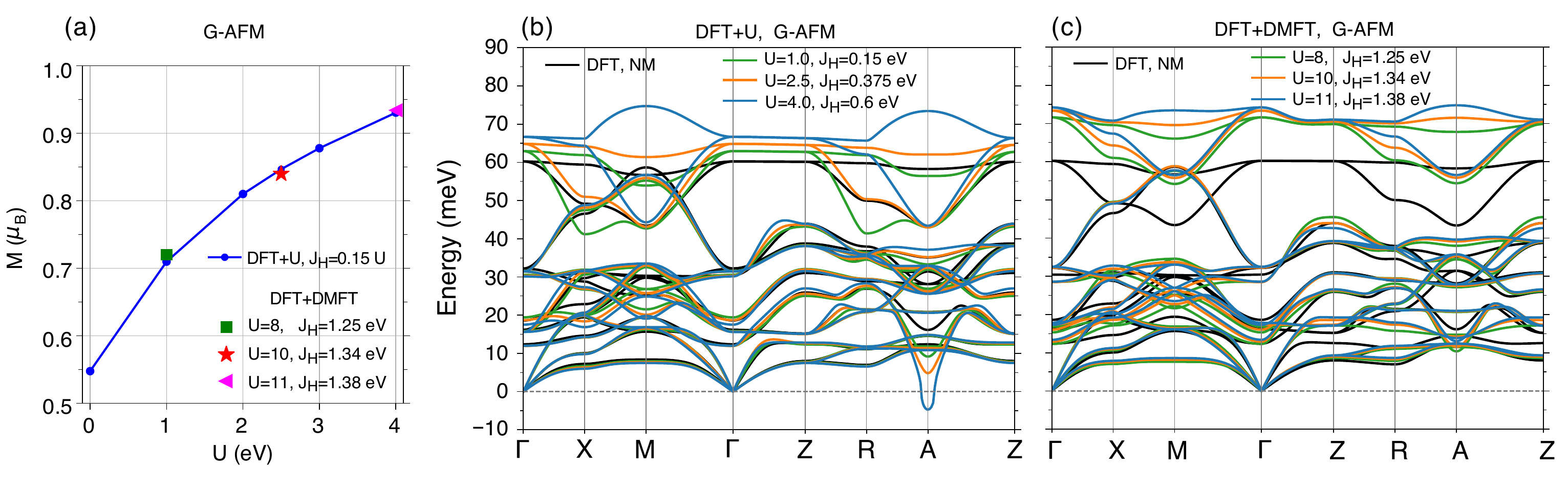}
    \caption{(a) Comparison of ordered magnetic moments calculated by DFT+$U$ and DFT+DMFT in G-AFM state, where the blue curve is for DFT+$U$, the green rectangle, red star and magenta triangle are for DFT+DMFT. Phonon spectra of LaNiO$_2$ in the G-AFM state calculated by (b) DFT+$U$ and (c) DFT+DMFT methods, respectively. The phonon spectra have been unfolded into the $1\times 1\times 1$ non-magnetic cell.}
    \label{fig:dftu_phonon}
\end{figure*}

\begin{figure}
    \centering
    \includegraphics[width=0.48\textwidth]{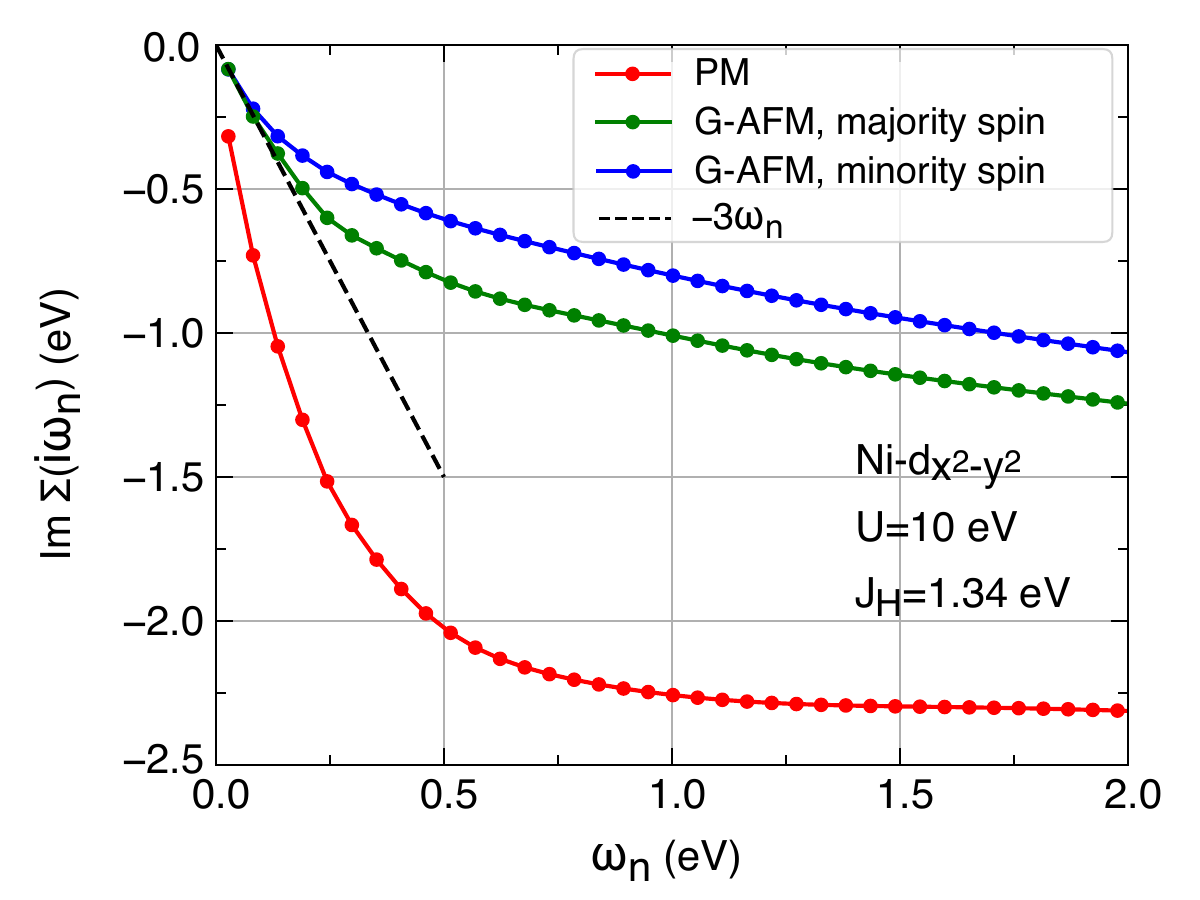}
    \caption{The imaginary part of self-energy at Matsubara frequency Im$\Sigma(i\omega_n)$ for Ni-$d_{x^2-y^2}$ orbital, calculated by DFT+DMFT for PM and G-AFM state of LaNiO$_2$, respectively. $U=10$ eV, $J_H=1.34$ eV, $T=100$ K. The black dashed line indicates a slope of $-3$.}
    \label{fig:sigma}
\end{figure}

\begin{figure*}
    \centering
    \includegraphics[width=1.0\textwidth]{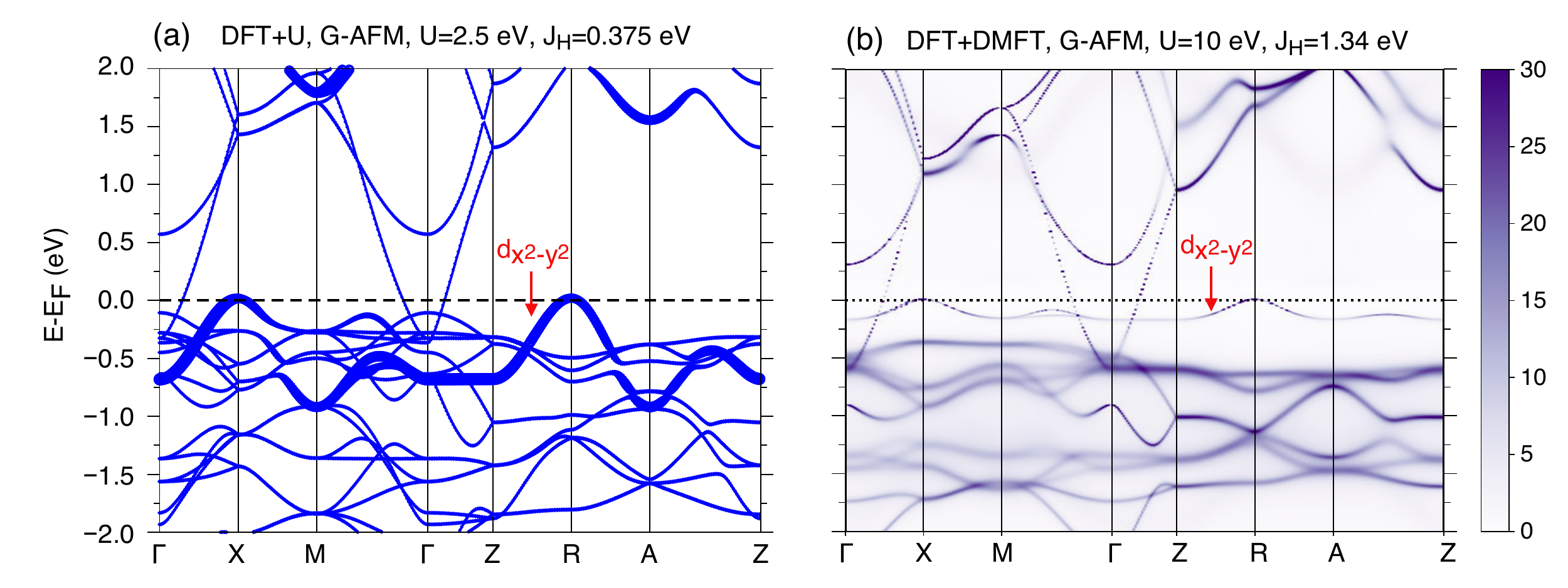}
    \caption{Comparison of the band structures of LaNiO$_2$ in G-AFM state calculated by (a) DFT+U at $U=2.5$ eV, $J_{\text{H}}=0.375$ eV and (b) DFT+DMFT at $U=10$ eV, $J_{\text{H}}=1.34$ eV. The $d_{x^2-y^2}$ band is shown as fat band in (a). It should be noted that the spin-polarization strength is the same between (a) and (b). Here, the band structures are plotted in terms of the AFM BZ.}
    \label{fig:dftu_band}
\end{figure*}

\begin{figure}
    \centering
    \includegraphics[width=0.5\textwidth]{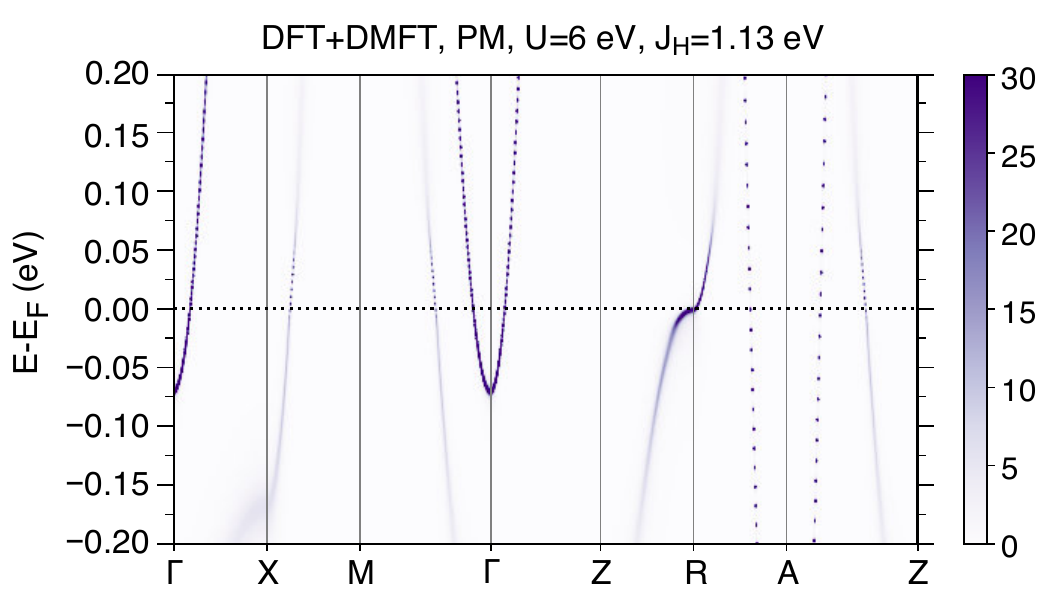}
    \caption{The band structure of LaNiO$_2$ calculated by DFT+DMFT at $U=6$ eV, $J_{\text{H}}=1.13$ eV in the PM state. No obvious kinks can be identified around $0\sim -150$ meV.}
    \label{fig:akw_U6}
\end{figure}

\section{Results}
Fig.~\ref{fig:dmft_phonon}(a) shows the phonon spectra of LaNiO$_2$ calculated by DFT in the NM state (solid curves) and by DFT+DMFT in the PM state at $U=6$ eV (dashed curves), respectively. The red color highlights the optical Ni-O bond-stretching phonons~\cite{Reznik:2012,Reznik:2021}, i.e., vibrations of O atoms along the Ni-O bonds within the NiO$_2$ plane. The atomic vibrations of four such modes (indicated by the black arrows in Fig.~\ref{fig:dmft_phonon}(a)), including full-breathing (FB), half-breathing(HB), quadrupolar (Quad) and a stripe mode 3$a_0$ ($Q=(1/3, 0, 0))$ are illustrated in Figs.~\ref{fig:dmft_phonon}(d)-(g).  

The DFT calculated phonon spectra are consistent with previous studies~\cite{ChenHanghui:2022,Meier:2024,DengHuiXiong:2024,NieYuefeng:2024}. Overall, the two methods give very similar phonon spectra, except the two branches of Ni-O bond-stretching phonons (red color). In the NM state, the maximum phonon frequency is less than 63 meV. Going from NM to PM state, these two branches  harden significantly. Increasing Hubbard $U$ will make them even harder (see Fig.~\ref{fig:dmft_phonon}(b)), such that the maximum frequency is larger than 70 meV. This is consistent with the onset energy of the kink (70$\sim$100 meV) observed in the electronic structures of LaNiO$_2$ by the ARPES experiment~\cite{NieYuefeng:2024}, indicating possible strong electron coupling to the Ni-O bond-stretching phonons. 

As shown in Fig.~\ref{fig:dmft_phonon}(c), the overall phonon frequencies decrease by Sr-doping. However, the hardening effect of the Ni-O bond-stretching phonons still persists when going from NM to PM states. We speculate that such hardening effects should be universal in the family of the infinite-layer nickelates \textit{R}NiO$_2$, regardless of the species in $R$-site.

Previous study has found that fluctuating local magnetic moments will significantly change the phonon frequencies in bulk FeSe without long-range magnetic order~\cite{KristjanHaule:2020}. Here, we will demonstrate that they are also the key factors in inducing the phonon hardening effect in LaNiO$_2$. For this purpose, we go to the limit with static magnetic moments in a magnetically ordered state. Although no long-range magnetic order has been found in LaNiO$_2$ by the available experiments so far~\cite{Hayward:1999}, stable magnetic orders can be still obtained by static mean-field calculations such as DFT and DFT+$U$ or dynamical mean-field calculations at very large $U$ ($>$ 7 eV)~\cite{Pascut2023}. Previous calculations show that the C-type and G-type antiferromagnetic configurations are competing ground states with very close energies~\cite{YilinWang:2020,LiuZhao:2020,Leonov:2020,Ryee:2020,Lechermann:2021mag,WanXiangang:2021}. Here, we will focus on the G-AFM state (see Fig. S1(d)~\cite{suppl}). 

Fig.~\ref{fig:dftu_phonon}(a) shows the ordered magnetic moments in G-AFM state calculated by DFT+$U$ (blue curve) and DFT+DMFT (green square, red star and magenta triangle) as functions of Hubbard $U$ and Hund's coupling $J_H$. In DFT+$U$, the magnetic moment increases from 0.55 $\mu_{B}$ to 0.93 $\mu_{B}$ when $U$ increases from 0 to 4 eV (with $J_H=0.15U$). In sharp contrast to DFT+$U$, magnetic moments of comparable magnitudes are obtained at much larger $U$ and $J_H$ in DFT+DMFT. For example, magnetic moments of about 0.72, 0.84, 0.93 $\mu_B$ obtained in DFT+DMFT at $U=8, 10, 11$ eV, which correspond to those obtained in DFT+$U$ at $U=1, 2.5, 4$ eV, respectively. It should be noted that both DFT+$U$ and DFT+DMFT calculations are performed in WIEN2K code, using similar local orbitals to define the correlated subspace. We also note that performing DFT+DMFT calculations with a Hartree-Fock impurity solver will yield the same results as DFT+$U$~\cite{kotliar:2006}. Therefore, here the difference in Hubbard $U$ for obtaining the same magnetic moments between DFT+$U$ and DFT+DMFT should be mainly attributed to the CTQMC impurity solver that can exactly solve the Anderson impurity model and capture all the local dynamical correlations. The local dynamical effects lead to strong fluctuation of local magnetic moments, which results in much larger $U$ for magnetic ordering.

We then compare the calculated phonon spectra of LaNiO$_2$ in G-AFM state between DFT+$U$ and DFT+DMFT methods at the same strength of spin polarization, i.e. $U=$1.0 eV (8 eV), 2.5 eV (10 eV) and 4.0 eV (11 eV) in DFT+$U$ (DFT+DMFT). The results are shown in Figs.~\ref{fig:dftu_phonon}(b) and (c), respectively. Overall, both methods yield hardening of the Ni-O bond-stretching phonons when going from the NM to G-AFM state, and increasing $U$ will make them even harder. This indeed points to the crucial roles of the magnetic moments in the lattice dynamics and confirm the DFT+DMFT results in PM state. However, there are several differences in the hardening trends between DFT+$U$ and DFT+DMFT. The hardening effect in DFT+$U$ is less rigid than that in DFT+DMFT, comparing to the phonon dispersion in NM state. The overall hardening strength in DFT+$U$ is slightly weaker than that in DFT+DMFT, in particular along the $\Gamma$-X direction. An exception is around $M$-point where DFT+$U$ with $U\ge2.5$ eV gives comparable hardening strength as DFT+DMFT with $U=11$ eV. We also note that there is one branch of phonon that becomes imaginary at $A$-point in DFT+$U$ at $U=4$ eV, $J_H=0.6$ eV. This corresponds to an out-of-phase rotation of the NiO$_4$ squares around the $c$-axis, which was also discussed in previous DFT or DFT+$U$ studies of other infinite-layer nickelates~\cite{ChenHanghui:2022,Philippe:2023,DengHuiXiong:2024,Cano:2022,Julien:2022}. However, it does not become imaginary in DFT+DMFT at $U=11$ eV with the same strength of spin-polarization as DFT+$U$ at $U=4$ eV. These differences should be attributed to the local dynamical correlation effects that are absent in DFT+$U$ method.

The local dynamical correlation effects are described by the frequency-dependent self-energy. As shown in Fig.~\ref{fig:sigma}, the self-energy of Ni-$d_{x^2-y^2}$ orbital indicates that, although the electron correlation strength significantly decreases when going from PM to G-AFM state, the G-AFM state still shows pronounced dynamical electron correlation, which is induced by the strong local magnetic fluctuation. As guided by the black dashed line in Fig.~\ref{fig:sigma}, the slope of the imaginary part of self-energy at low-frequencies is about $-3$, which gives a quasi-particle renormalization factor $Z=1/4$ for $d_{x^2-y^2}$ orbital, according to $1/Z=1-\frac{\partial \text{Im}\Sigma(i\omega_n)}{\partial \omega_n}|_{\omega_n \rightarrow 0}$. This will induce pronounced renormalization of $d_{x^2-y^2}$ bands even in the AFM state, which is demonstrated in Fig.~\ref{fig:dftu_band}. The overall band structures between DFT+$U$ at $U=2.5$ eV and DFT+DMFT at $U=10$ eV are similar, but the $d_{x^2-y^2}$ band near the Fermi level becomes much narrower in DFT+DMFT than that in DFT+$U$. 

These results suggest that, for materials like infinite-layer nickelates in which there is no long-range magnetic order but strong local magnetic fluctuations, the local dynamical correlations and magnetic fluctuation effects should be fully taken into account when studying their lattice dynamics. The DFT+DMFT method can accurately and directly describe such effects in the PM state. Approximating the spins as disordered in a large supercell within DFT codes~\cite{Alling:2016} could be another way to describe such physics, for example, good results have been obtained for the phonons of Pu~\cite{Soderlind2015}.

The phonon hardening indicates a strong interplay between the Ni-O bond-stretching vibrations and the electron correlation of Ni-$3d$ orbitals. To confirm this, we perform DFT+DMFT calculations in PM state to compare the correlation strength of four distorted structures with frozen Ni-O bond-stretching phonons, i.e., FB, HB, Quad and 3$a_0$ (Figs.~\ref{fig:dmft_phonon}(d)-(g)), to the pristine structure. The displacements of O atoms are fixed to be 0.05 \AA\: in the distorted structures. In LaNiO$_2$, the most strongly correlated orbital is Ni-$3d_{x^2-y^2}$, so we compare the quasi-particle mass-enhancement $m^{*}/m^{\text{DFT}}$ of $d_{x^2-y^2}$ orbital at $U=6$ eV. Larger mass-enhancement indicates stronger electron correlation. As shown in Table~\ref{tab:mass}, the electron correlations are indeed very sensitive to the Ni-O bond-stretching distortions. The correlation strengths of both Ni1 and Ni2 sites significantly decrease for the FB, Quad and 3$a_0$ structures. For example, it decreases by 29\% and 18\% for Ni1 and Ni2, respectively, in the FB structure. For the HB structure, the correlation strength of Ni2 site slightly increases but significantly decreases for the Ni1 site. 

Our DFT+DMFT calculations give a phonon frequency of about 70 meV in the reasonable Hubbard $U$ regime for LaNiO$_2$. It is consistent with the kink observed in its electronic structures by ARPES experiment~\cite{NieYuefeng:2024}. It should be noted that pure electron correlations will also induce such kinks~\cite{Byczuk2007}. As shown in Fig.~\ref{fig:akw_U6}, we plot the DFT+DMFT calculated spectral function at $U=6$ eV and $J_H=1.13$ eV in PM state. No obvious kinks can be identified around 0 $\sim -150$ meV. We note that the DFT+DMFT calculation can well produce the band renormalization measured by ARPES experiment at even smaller Hubbard $U$ ($\sim 4$ eV)~\cite{NieYuefeng:2024}, so the kink cannot be induced by electron correlations. This, together with the very sensitive electron correlations to the Ni-O bond-stretching phonons, provide evidence to ascribe the kink to possible strong electron coupling to the Ni-O bond-stretching phonons. Their importance to superconductivity deserves further studies. A recent work based on GW calculations~\cite{Steven:2024} found the important roles of the electron-phonon couplings to the superconductivity in Nd$_{1-x}$Sr$_x$NiO$_2$.

The phonon hardening suggests that strong electron correlation will go against a static lattice distortion of Ni-O bond-stretching. To confirm this, we use DFT+DMFT to perform atomic relaxation for four distorted structures of LaNiO$_2$ and La$_{0.5}$Sr$_{0.5}$NiO$_2$ with finite displacements of O atoms illustrated in Figs.~\ref{fig:dmft_phonon}(d)-(g). The relaxation are performed in the PM states at $U=6$ eV and temperature down to 58 K. We find that all the relaxations eventually converge to the pristine structure without any distortions. We also perform fully charge-self-consistent DFT+DMFT calculations by imposing finite charge disproportionation among Ni-sites in the first iteration. However, they also converge to the solutions with uniform charge distribution. Therefore, our results suggest that, at least, the strong \textit{local} electron correlation will not favor charge orders such as $3a_0$~\cite{ZhouKejin:2022,Preziosi:2022,LeeWeisheng:2022,Rossi:2024,ShenKM:2024} if they couple to the Ni-O bond-stretching phonons. It should be noted that possible non-local Coulomb interaction has not been explicitly included in our present DFT+DMFT calculations, so we cannot exclude the possibility of CDW instability induced by non-local correlation effects. Since this is not the main scope of the present work, we defer this to future study. 

\begin{table}
    \centering
    \caption{Quasi-particle mass-enhancement $m^\star/m^{\text{DFT}}$ of the Ni-$d_{x^2-y^2}$ orbital calculated by DFT+DMFT at $U=6$ eV, $J_H=1.13$ eV and $T=100$ K, for pristine and four distorted structures. The displacements of O atoms in the distorted structures are 0.05 \AA. For FB, HB and $3a_0$, the two numbers are for Ni1 and Ni2 sites, respectively.}
    \begin{ruledtabular}
    \begin{tabular}{cccccc}

                   & pristine & FB & HB & Quad & $3a_0$\\
         \hline
         $m^\star/m^{\text{DFT}}$  & 4.60 & 3.26/3.77 & 3.62/4.87 & 3.99 & 3.75/4.37 \\
    \end{tabular}
\end{ruledtabular}
    \label{tab:mass}
\end{table}

\section{Conclusion} 
To summarize, we show that the dynamical electron correlation and fluctuating local magnetic moments in the PM state will induce significant hardening of the Ni-O bond-stretching phonons in LaNiO$_2$. Our results emphasize that the effects of local dynamical correlations and magnetic fluctuations should be fully taken into account when describing the lattice dynamics of the infinite-layer nickelates without long-range magnetic orders.  We show that the strong \textit{local} Coulomb interaction will not favor charge orders that couple to the Ni-O bond-stretching phonons in LaNiO$_2$. This supports the recent experimental finding that the $3a_0$ charge order is absent in the pure LaNiO$_2$ films without extra oxygen sites~\cite{ShenKM:2024}. Our results also provide evidence to ascribe the kink observed in recent ARPES experiment to possible strong electron coupling to the Ni-O bond-stretching phonons. 

\section{Acknowledgement} 
This work was supported by the National Key R\&D Program of China (Grant No. 2023YFA1406304), the National Natural Science Foundation of China (No. 12174365), the Innovation Program for Quantum Science and Technology (No. 2021ZD0302800) and the New Cornerstone Science Foundation. All the calculations were performed in Hefei Advanced Computing Center, China.


\bibliography{main}

\end{document}